\documentclass[12pt,letterpaper]{article}
\usepackage{amsmath,amssymb,pgf,pgfarrows,pgfnodes,float,appendix, hyperref}
\usepackage{graphicx}
\usepackage{subfigure}
\usepackage[margin=0.9in]{geometry}

\newcommand{\be}{\begin{equation}}
\newcommand{\ee}{\end{equation}}
\newcommand{\bea}{\begin{eqnarray}}
\newcommand{\eea}{\end{eqnarray}}

\title{{\rm\footnotesize \qquad \qquad \qquad \qquad \qquad \ \qquad \qquad \qquad \ \ \ \ \ \                  RUNHETC-2020-03}\vskip.5in   Finite Deformations of Quantum Mechanics  }
\author{Tom Banks\\
Department of Physics and NHETC\\
Rutgers University, Piscataway, NJ 08854\\
E-mail: \href{mailto:banks@physics.rutgers.edu}{banks@physics.rutgers.edu}
\\
\\
}
\date{}
\begin{document}
\maketitle

\begin{abstract}
We investigate modifications of quantum mechanics (QM) that replace the unitary group in a finite dimensional Hilbert space with a finite group and determine the minimal sequence of subgroups necessary to approximate QM arbitrarily closely for general choices of Hamiltonian.  This mathematical study reveals novel insights about 't Hooft's Ontological Quantum Mechanics, and the derivation of statistical mechanics from quantum mechanics.  We show that Kornyak's proposal to understand QM as classical dynamics on a Hilbert space of one dimension higher than that describing the universe, supplemented by a choice of the value of a naturally conserved quantum operator in that classical evolution, can probably be a model of the world we observe.
\end{abstract}

\section{Introduction}

It is notoriously difficult to modify quantum mechanics.  Non-linear modifications or violations of unitary evolution of density matrices lead to dramatic contradictions with experiment, unless the coefficients of the modified terms are made extraordinarily small\cite{bpsp}.  In this paper we will explore a natural modification of QM, which can also approximate its results with arbitrary precision, but preserves the attractive features of the original theory.

The mathematical definition of infinity and the inevitable imprecision of measurement apparatus at any given time imply that it is sufficient to study finite dimensional quantum systems.  In a classical system with the same number of states, this would be sufficient to render the system completely finite.  Time evolution would be a discrete permutation of the states.  In quantum mechanics even a two state system has a continuous infinity of incompatible observables and continuous time evolution.  The question that we will address in this paper is whether one can replace the axioms of QM with those for a discrete system, which can approximate QM with any required degree of accuracy.

There are several obvious ideas, which do not work.  We could replace the complex number field by the complex rationals.  This is a cheat, because rationals are dense in the real numbers and all we really mean by the real numbers is limits of sequences of rationals.  Choosing finite fields\cite{finitefield} does not appear to work because vector spaces over finite fields do not have scalar products.  The authors of\cite{finitefield} provide definitions of probability for these vector spaces, but it is not clear whether a consistent formalism including probability conserving time evolution exists or whether their are limits where that formalism can approximate the empirical successes of QM.

 Instead we will ask whether we can restrict the group of unitaries to a finite group in a sensible way.  Some might consider this to be no modification of QM at all, but there is still an interesting question of how well such a finiteness restriction can reproduce experiment. Quantum evolution is defined by a choice of an initial basis, $| e^i_{in} \rangle$ and a sequence of unitaries $U(t,t_0)$ for discrete times $t$ such that
\begin{equation} U(t_0, t_0) = 1 \end{equation} and \begin{equation} U(t,t_0) = U(t,s) U(s,t_0) , \end{equation} for any $t < s < t_0$.  Alternatively, we could give up the constraint on $U(t_0, t_0)$ and choose a fixed initial basis.   The question we want to ask is: what is the smallest discrete group that can approximate {\it any} continuous unitary evolution with some required accuracy? Note that the word {\it any} is important.  If we restrict attention to a single time independent system, then it is clear that restricting to the group $Z_k^N$ where $N$ is the dimension of the Hilbert space,  and the initial basis is chosen to be the basis of Hamiltonian eigenstates, can do the job.  The accuracy of approximation will increase with $k$ and will depend on the eigenvalues of the Hamiltonian.   However, we want to make a choice of group that will approximate any system.

Our question has relevance to the attempt by 't Hooft\cite{tHqm} to replace quantum mechanics with a classical evolution of some particular basis.   That is, one postulates that what is "really" going on in the universe is permutation of some particular, {\it ontological} basis.  We will call such a restriction Ontological Quantum Mechanics (OQM). The states we "observe" are supposed to correspond to superpositions of these basis states.  One way of explaining this is to accept all possible operators as physical and claim that the measurement apparatus available to us does not commute with the projectors on ontological basis states.  As 't Hooft points out\cite{2019}, this leads to "ontological conservation laws":  The initial probabilites of being in each ontological basis state are conserved in time.   This restricts the extent to which such a system can mimic ordinary QM.

't Hooft argues instead that collective coordinates of macroscopic objects are linear combinations of the microscopic ontological projectors, avoiding the usual decoherence arguments, which are invoked to explain the fact that collective coordinates appear to obey classical statistical equations, which are linear in probabilities\footnote{The present author finds this part of 't Hooft's program difficult to understand.
.  }. He attributes the appearance of quantum mechanics, at scales between that of the standard model of particle physics and the macroscopic scale, to a renormalization group transformation based on cutting off the quantum mechanical spectrum of the ontological evolution operator.  Since the evolution operator is diagonal in the Fourier transform of the ontological basis, this cutoff introduces superpositions of ontological states.  We will show below that this procedure is analogous to introducing a non-gauge invariant cutoff in a gauge theory.  We'll argue both that there's an alternative, gauge invariant, procedure for coarse graining in time, and that for a time independent evolution operator, the insistence on determinism for the microscopic dynamics, restricts one to permutations with an equally spaced spectrum.  We propose that a cosmological version of 't Hooft's program might solve this problem.

Another area of current research that is illuminated by our study is the arena of computational complexity.  There is a lot of evidence that the eigenvalue statistics of chaotic quantum systems of large dimension closely match those of random ensembles of Hamiltonians\cite{shenkeretalandrefs}.  There appears to be an assumption in the field that long time evolution of such a system will in fact be random, eventually approaching any state in a finite dimensional Hilbert space with arbitrary precision if the evolution is restricted to a finite dimensional subspace. A simple corollary to our results is that this is not true.  Evolution by a fixed Hamiltonian is even more restrictive than evolution under a fixed finite subgroup of $SU(N)$.  The "ontological conservation laws" for this case are just the initial probabilities that the system was in each of its energy eigenstates.  Thus, one cannot reach a state with one value of these probabilities from one with a different value, by Hamiltonian evolution.  There's a similar problem with periodic time dependent evolution if the period is shorter than the quantum recurrence time of the system.   Note by the way that this no-go statement does not forbid quantum recurrences.  The initial state and the recurrent state will have the same values of all the energy probabilities.

The finite form of quantum mechanics discussed here was invented in\cite{finite}.  That paper did not investigate the extent to which such a modification of the quantum theory was consistent with the successful modeling of real systems, but it made the crucial observation that restriction of allowed operators to the large representation of $S_N$ on $N$ dimensional space, was enough to produce amplitudes with quantum interference.
In this paper we will show that, for $N$ large enough to account for the observable universe, there is no problem in reproducing all possible experiments accounted for by any conventional quantum theory, as long as the period of time over which experiments can be done is much less than $N$ in fundamental units.  The most likely reason for such a restriction in the real world, is that our universe asymptotically approaches de Sitter space, that is either stable or has a decay time much longer than the de Sitter Hubble time $T_H$.  If we accept the Covariant Entropy Principle\cite{bhgthjfsb} then $N = e^{\pi T_H^2}$, when $T_H$ is expressed in Planck units.  $S_N$ evolution allows for quantum energy differences of the form $2\pi \sum_i \frac{k_i}{p_i}$, where $p_i$ is any set of primes satisfying $\sum p_i = N$ and $0 \leq k_i < p_i$.  Evolution over time periods comparable to $T_H^p$ is sensitive only to differences of order $T_H^{-p}$.  For large enough $p$ this is larger than the lifetime of any localized object in dS space. Thus, any Hamiltonian whose dynamics is detectable by local experiments in an asymptotically dS universe is easily mimicked by $S_N$ evolution.  The spectrum of the permutation operator is sufficiently chaotic to avoid apparent integrable behavior, which could contradict experiment.  However, we'll see that for a time independent system, the fundamental hypothesis of determinism restricts the accessible energy spectrum to that of a single cyclic permutation, which is an integrable rather than a chaotic spectrum.

This paper mixes together several different kinds of questions.  There are purely mathematical questions of the extent to which finite subgroups of $SU(N)$ can approximate generic unitary evolution and/or the special forms of evolution operators appearing in known models.  There is the somewhat different question of the extent to which the approximation can be mathematically inaccurate but sufficient to reproduce the success of quantum mechanics in the real world.  Then there is the specific question of whether 't Hooft's program can approximate the real world.  Finally there is the philosophical problem of whether determinism is a logically necessary stance.  The consideration of philosophical problems is relegated to an appendix, which should nonetheless be read.  The author has found it impossible to completely separate the other sorts of questions from each other.  They are tied together by the same mathematics.  Readers should be aware that some of the arguments go back and forth between them.

\section{Finite Subgroups of $SU(N)$} 

The most general finite subgroup of $SU(N)$ for large enough $N$ is a semi-direct product of $S_{N + 1}$ and a finite abelian group\cite{finiterefs}.  To see that $S_{N+1}$ can act on $N$ dimensional Hilbert space, note that it has an obvious action on a Hilbert space of dimension $N + 1$, as the permutations of a particular orthonormal basis $| e_i \rangle$.   This action leaves the vector $\sum_i | e_i \rangle$ invariant and so acts on an $N$ dimensional subspace.  Note that the vast majority of linearly independent operators in $N + 1$ dimensions, with $N \gg 1$ do not make transitions between the two irreps.

 $S_N$, acting on a fixed basis, has a $Z_N$ subgroup generated by the familiar "shift" operator $V$, satisfying $V^N = 1$. $S_{N + 1}$ is obtained by adding one more generator to $S_N$,  the Fourier transform operator \begin{equation} F = \sum_i | u_i \rangle\langle v_i | , \end{equation} which maps the original basis of $U$ eigenstates (which is the ontological basis for the canonical $S_N$ subgroup of $S_{N + 1}$), into the basis of $V$ eigenstates.  Larger finite subgroups of $U(N)$ are of the form ${\cal G}_{\cal F} (N) = {\cal F}^{2N} \ltimes SU(N + 1)$, where ${\cal F}$ is a finite abelian group.   Here each individual ${\cal F}$ is a group of phases acting on a single element of either the $| u_i \rangle$ or $| v_i \rangle$ bases. The semi-direct product transforms this single element phase into independent ${\cal F}$ phases on each of the $ | u_i \rangle$ and $ | v_i \rangle$ basis vectors as well as permutations of each basis and the Fourier transform between them.  All of these groups leave invariant
\begin{equation}{\cal S}_k \equiv \sum_i  ( p^k(u_i) ) + p^k(v_i) ) , \end{equation} which divide the space of states into discrete equivalence classes.   If the initial state is one of the $| u_i \rangle$ or $| v_i \rangle$ basis vectors then
 \begin{equation} {\cal S}_k = 2 , \end{equation} for all $k$.  For generic unit vectors, all the probabilities $p(u_i)$ and $p(v_i)$ are less than $1$ so all of ${\cal S}_k$ go to zero exponentially for large $k$.  Indeed, in the large $k$ limit, ${\cal S}_k $ is dominated by the largest of these $2N$ probabilities.
 Thus, as $N$ gets large it gets harder and harder to approximate a generic unitary.  
 
 On the other hand, the Hamiltonians with which we model just about every physical system known to man, as well as many that belong to the realm of science fiction, have the form  $H = h_u (U) + h_v (V)$\footnote{The prominent exception is the Hamiltonian for particles in an external magnetic field. However, these also have a Feynman path integral formulation, so our remarks apply to them.}. For such Hamiltonians, the Trotter product formula tells us that if we approximate continuous time evolution by evolution over short enough discrete intervals then the short time evolution operators are the product of a function of $U$ and a function of $V$.  These can all be approximated with arbitrary precision by an element of the $ {\cal F}^N (U) \times {\cal F}^N (V)$ subgroup of ${\cal G}_{\cal F} (N)$ for a large enough finite abelian group. 
 
 Of course, for any single system, with a time independent Hamiltonian, there is always a basis in which the evolution operator is well approximated by an element of ${\cal F} (U)$ for large enough finite abelian group.  Our point is that the validity of the Trotter product formula for a wide class of known Hamiltonians shows that we can reproduce many different systems with the same discrete subgroup of $SU(N)$.  More specifically, given any Hamiltonian $H$, and $U(1)^{N - 1}$ the Cartan torus that commutes with it, we define the {\it Trotter (T) equivalence class of the Hamiltonian} to be all those Hamiltonians $H + \Delta$ for which the short time evolution operator can be approximated by an operator of the form $f(U) g(V)$, where $U$ is the clock operator in the Cartan torus, $V$ its canonical shift operator, and $f$ and $g$ unitary operators that commute with $U$ and $V$ respectively .   Then any class of quantum systems with (possibly) time dependent Hamiltonians all belonging to the same T-equivalence class can be replaced by one for which quantum evolution belongs to ${\cal G}_{\cal F} (N)$, with an accuracy that increases with the size of ${\cal F}$.   This is simply the statement that $f$ and $g$ are diagonal matrix elements of phases, representations of ${\cal F}$ are one dimensional, and can approximate any phase as the size of ${\cal F}$ increases.  
 
 In fact, we can restrict attention to the subgroup of ${\cal G}_{\cal F} (N)$ generated by ${\cal F} (U)$ and the Fourier transform to approximate any system in a given T-equivalence class.  The class of systems well approximated by evolution in ${\cal G}_{\cal F} (N)$ with a given choice of clock operator $U$ is thus larger than a T-equivalence class, but I have not found a simple alternative characterization of it.  One should note in particular that for ${\cal F}$ containing elements with order greater than the maximal order of an element of $S_{N + 1}$ (which grows like $e^{c\sqrt{N {\rm ln}\ N}}$ )the more general ${\cal G}_{\cal F} (N)$ evolution will have longer periods than any evolution in the subgroup.  
 
  \section{Ontological Quantum Mechanics}
  
  We will use the phrase $OQM_N$ to mean a quantum system in $N$ dimensional Hilbert space with unitary group restricted to its $S_N$ subgroup, acting on a special ontological or  "ontic" basis.  There are actually two possible versions of this theory, depending on whether one treats the $U(1)^N$ subgroup diagonal in the ontic basis as a set of physical transformations, or as gauge transformations.   If they are gauge transformations, then operators that are not diagonal in the ontic basis are considered unphysical. This is a mathematical definition of what we mean by a deterministic classical system.  If off diagonal operators are physical, then the theory has all of the "peculiar" properties of ordinary QM.  In particular, the transition probabilities between eigenstates of any off diagonal Hermitian operators do not satisfy the Probability Sum Rule (PSR):
  \begin{itemize}
  \item The probability to go between eigenvalues $A_i$ and $A_f$ in time $t$ is the sum of the probabilities for all intermediate histories $A(t)$ that go between $A_i$ and $_f$ in time $t$.
  \end{itemize}
  This sum rule is the basis of Bayes' conditional probability rule, which tells us how to verify a probabilistic theory in actual experiments.   All of the "weirdness" of quantum mechanics stems from its violation of this rule.  The rule follows in any statistical theory in which probabilities, rather than probability amplitudes, satisfy linear evolution equations.
  
  In the gauged version of the theory, superpositions of ontic states are unphysical (though they may be used for mathematical convenience in doing calculations).  We will call this version of the theory, the one adopted by 't Hooft, $DOQM_N$, where the $D$ stands for determinism.  See the Appendix for a discussion of determinism.  Mathematically, a deterministic theory is one in which the group of unitaries is restricted to $S_N$ {\it and} $U^N (1)$ is a gauge symmetry.
  
  $DOQM_N$ is a discrete version of Koopman's formulation of classical mechanics in Hilbert space\cite{koopman}. This is often called "pre-quantization" in the mathematical theory of geometric quantization.  Wave functions are complex functions on phase space (actually, sections of a line bundle over phase space) and the Hamiltonian is
  \begin{equation} H = i {\cal L} = i(\partial_q E \partial_p - \partial_p E \partial_q) . \end{equation}   It is easy to see that the diagonal density operator in the $p,q$ basis satisfies the classical Liouville equation and that delta function densities propagate to delta functions at points satisfying the usual Hamilton equations of motion.  Operators diagonal in the $p,q$ basis transform into other diagonal operators and we can impose a gauge invariance $\psi \rightarrow e^{i\theta (p,q)} \psi$, which renders non-diagonal operators unphysical.
  This formalism is easily generalized to Hamiltonian systems on any symplectic manifold.
  
  An important feature of Koopman's dynamics is that time translation invariance leads to two different operator conservation laws.  Both $H$, and the diagonal operator $E(p,q)$, the classical energy, are conserved.  In the gauged version of the theory, the gauge invariant states are states in which $H$ is infinitely uncertain, and only the classical energy $E$ is a useful constraint on the dynamics.  The analog of $H$ in $DOQM_N$ is the time independent permutation operator $S$, for evolution over the elementary time interval.   Like every permutation it is a product of commuting cycles $S = C_1 \ldots C_n$ of prime length $p_n$.  The mutual eigenvales of these operators are the analogs of the eigenvalues of the Hamiltonian $H$, while the cycle lengths $p_i$ are the analogs of the classical energy $E$.  In the ontic states, the only physical states of $DOQM_N$, the eigenvalues of the $C_l$ are maximally uncertain,  there is equal probability to have any collection of these eigenvalues.  The $p_l$ are completely certain in the ontic states.  They are physical observables in $DOCM$.  Despite their global sounding definition, they can be computed by a straightforward but tedious process, from the matrix representation of $S$ in the ontic basis, as functions of the "point in phase space".  That is, every ontic state evolves under a given permutation, through a cycle of some length $p_l$.  The formula assigning $p_l$ to each point is the analog of the formula for the Hamiltonian as a function on phase space in classical mechanics.  Note that the lengths $p_l$ are the things that determine the notions of slow and fast motion of the ontic variables.  In order to have anything slowly varying on the fundamental time scale of this discrete QM, we must have long cycle lengths.  Slowly varying physical degrees of freedom are operators of the form 
  \begin{equation} A = \sum_{i \in C_l} a(i) | o_i \rangle \langle o_i | , \end{equation} where $a(i)$ varies on a scale $1 \ll \Delta i \ll p_l $.  The present author's understanding of "the renormalization group" for such a system would be a systematic derivation of closed stochastic equations for variations of these slow variables on time scales of order $\Delta i$ , whose only reference to the underlying microsystem was a probability distribution for initial conditions for combinations of ontic projectors orthogonal (in the sense of the trace inner product on operators) to the slow variables.  The equations are stochastic because they have no place to input the initial conditions for the variables orthogonal to the slow ones.  The probabilities for slow variables would obey the PSR.   Note that the classical RG defined by this procedure has an emergent notion of "distance on phase space" associated with it.  Points in phase space are close to each other, by definition, if they can be reached by time evolution over intervals $< \Delta_i$.
  
  Koopman's quantum classical mechanics gives us many explicit examples of such a classical RG procedure.  The simplest is a pair of coupled oscillators with frequencies $\omega \ll \Omega$.   We take $\Omega^{-1} \ll \Delta t \ll \omega^{-1}$ and solve the equations of motion for the fast oscillator with fixed slowly varying trajectory for the slow one.  We obtain a non-local effective action for the slow oscillator, which has a systematic derivative expansion controlled by powers of $\Omega^{-1} \Delta t $.  The action depends on the choice of initial conditions for the fast oscillator and we get a stochastic equation by integrating over them with some probability distribution.   It is quite clear that such a procedure will never generate predictions that violate the PSR.
  
  't Hooft's version of the RG for $DOCM_N$ is different.   Despite the fact that the ontic states have maximally uncertain values of the $C_l$ eigenvalues he implements the RG by integrating out the "high energy" eigenvalues of the $C_l$.   This of course produces "effective states" which are no longer ontic, and which of course exhibit quantum interference phenomena.  In the language of the paragraphs above, this is analogous to using a non-gauge invariant regulator to regulate a gauge theory.  Kenneth Wilson taught us that we could do this at the expense of adding "counterterms" to the effective theory to cancel all relevant and marginal operators that violate gauge invariance.  In the present context, it is not clear what the phrase "relevant" and "marginal" operators means.   't Hooft wants the violation of "deterministic ($U(1)^N$) gauge invariance" to account for the observable quantum properties of the standard model.  At the same time, he would like to claim that the values of collective observables of macroscopic bound states of standard model particles are simply functions of the underlying ontic variables.   The arguments that the latter claim is true are not expressed as mathematical equations that the present author currently understands. As Feynman's discussion of the double slit experiment illustrates, macroscopic measurements of things we identify as microscopic objects do not admit an interpretation in which the measured properties obey the PSR.  Our current theory of this, the standard model of particle physics, models the microscopic objects as elementary excitations of quantum fields, and the measuring apparatus itself as a collection of averages of composite fields over large volumes.  In this model, the measuring apparatus is not ontological, but obeys the PSR to extremely high accuracy because the macroscopic volume over which it is averaged is much larger than atomic length scales, and the apparatus is far from its ground state.  To carry out 't Hooft's program one would have to understand why his model of the apparatus as a function of the ontic variables, gave the same results as the quantum model of the apparatus.  
  
  We will go into more detail about determinism in the appendix, but it is worth while pointing out one more peculiar feature of 't Hooft's RG proposal.  In order to discuss the RG one has to restrict attention only to long cycles with $p_l \gg 1$.   In ordinary uses of the RG in quantum theory, we apply a cutoff on the values of the Hamiltonian, uniformly to all trajectories of the system.   However, since time independent $S_N$ quantum mechanics preserves individual cycles we have many choices about how to impose the cutoff.  The one that most resembles the conventional RG is to impose a cutoff $$ \sum_l \frac{k_l}{p_l}  < \phi / 2\pi , $$  where $\phi$ is the cutoff phase. This is completely bizarre in a deterministic theory.   In such a theory, our observations correspond to some particular cycle determined by the ontic initial condition at the beginning of our universe.  Why should we impose a cutoff whose value depends on what other alternative histories of the universe are doing?  This is true even if we posit that we must make only probabilistic predictions because we do not know some or all of the initial conditions of the system. The sensible thing for a classical physicist to do is to impose an energy cutoff on each individual trajectory, and then make predictions based on some probability distribution for which trajectory one took.
  
  Thus the alternative is to impose a cutoff on the individual $C_l$ eigenvalues, and then posit that our observations correspond to some particular cycle, with some initial probability distribution for which cycle it is.  The expression for a particular ontic state $| o_i \rangle$ belonging to a cycle $C_l$ in terms of the eigenstates of $C_l$ is
  \begin{equation} | o_i \rangle = \sum_{j=0}^{p_l - 1} e^{\frac{2\pi i j o_i }{p_l}} | j \rangle . \end{equation}
  If we cut off $j$ at some value $m$ and then take the overlap with $| o_i \rangle$ we get 
  \begin{equation} \langle o_i | o_k \rangle_m = \sum_{j = 0}^{m - 1}  
   e^{\frac{2\pi i j (o_k - o_i) }{p_l}} | j \rangle . \end{equation}
   If $m \sim p_l /2 \gg 1$ then this will be very small unless $o_k - o_i \ll p_l$.  Indeed, there are $m$ values of $k,k^{\prime}$ such that \begin{equation} \langle_m o_k |  o_{k^{\prime}}  \rangle_m \end{equation} is very close to zero, when $p_l$ is large.  We can use the Gram-Schmidt procedures to make these exactly orthogonal.  Let us use the notation $ | o_K \rangle_m$ for these orthogonalized basis states.
   
   The separation along the cycle between the centers of these coarse grained superpositions are of order $2$.   That is, the operator $C_l^2$ roughly moves each of these centers to the next one.  The quantum nature of the resulting dynamics consists in the mismatch \begin{equation}  | \Delta_K \rangle \equiv C_l^2 | o_K \rangle_m - | o_{K + 1} \rangle_m \end{equation}, between the unitary evolution through two time steps and evolution by a cyclic permutation of the $| o_K \rangle_m$ . $ | \Delta_K \rangle$ has a component in the subspace spanned by the $| o_K \rangle $ and a component orthogonal to it.  The spirit of the RG is to drop the orthogonal component and define an effective evolution in the "low energy" subspace, in the spirit of Brillouin-Wigner perturbation theory.  It is not entirely clear what the formula is for the effective vector $ | \Delta_K \rangle_{eff}$  in the $| o_K \rangle_m$ subspace.  Once this is sorted out one has defined an RG transformation that reduces the dimension of the Hilbert space and can proceed to iterate it.  The reduced dynamics will show quantum interference of histories if $| \Delta_K \rangle_{eff}$  is not small.  It is not clear to the present author how general a quantum dynamics can be generated by this procedure.
   
   The restriction of attention to fixed cycles also changes one's estimate of how well $S_N$ dynamics can mimic that of a general quantum system.  Fixing the cycle length means that we have $p_l$ states with "energies" $2\pi j/p_l$ for every $j$ between $0$ and $p_l - 1$.  
   
 Given a Hamiltonian with some natural energy scale $\epsilon$, in an $N$ dimensional space, we generally expect energy differences as small as $\epsilon/N$.  Observation of such small differences requires times of order $N \epsilon^{-1}$.  Even if we take $\epsilon = 10^{19} $ GeV, the Planck energy scale, the entire history of the universe takes only $10^{61}$ Planck times, whereas $N$ for even a piece of matter $.1$ cm. on a side is of order ${\rm Exp}\ ({10^{20}})$.  For the entire universe we probably have $N \sim {\rm Exp}\ ({10^{123}})$.  Thus cycles with $p_l \sim N$ are adequate for describing the dynamics of the universe for times of order its de Sitter Hubble time, as far as having both a large and small enough range of energy spacings.  However, the spectrum of a single cycle is evenly spaced, so it cannot describe the complex dynamics of the real universe.
 
 $S_N$ has elements with order $\sim e^{\sqrt{N {\rm ln}\ N}}$ because $\sum_l \frac{k_l}{p_l} $ can give a quite dense and chaotic spectrum\footnote{I do not know whether it is chaotic in the sense of random matrix theory.
}. This is because simultaneous periodic motion with periods satisfying $\sum p_l = N$ can have a common period as large as $p_1 p_2 \ldots p_n$.  However, in utilizing these very long mathematical periods we're again violating the spirit of deterministic dynamics.   In a deterministic  interpretation of our system, the universe is always in some ontic state.  Given a fixed evolution operator $S\in S_N$ for the elementary time interval, a choice of initial state determines a fixed cycle and the dynamics in the other cycles becomes irrelevant.  If $S$ contains multiple cycles, the system automatically has a conservation law for the projection on each of the cycles.  For fixed choice of those conservation laws the dynamics of the system is integrable, and cannot represent the universe we observe.

A possible way out of this impasse is to introduce time dependent dynamics, which violates all of these conservation laws.  This is certainly a reasonable hypothesis if one is trying to construct a cosmological theory, rather than {\it e.g.} reproduce the Standard Model of Particle Physics.  Two generic perturbations will have overlapping cycles, unless they commute and thus a succession of different permutations will violate all of the conservation laws above.  Time dependent ontological dynamics can actually execute all possible permutations and therefore give rise to the full spectrum of quantum energies, $\sum_l \frac{k_l}{p_l}$. In a time dependent system, it is therefore appropriate to discuss energy cutoffs of the full spectrum, rather than the individual cycles of a fixed permutation.

 In order to introduce the concept of quantum energy and to account for the observed approximate conservation of energy, we must make the time dependence $S(t)$ adiabatic during the period of the universe that we currently observe.  Thus, in the current era of the universe, the ontological variables are performing deterministic motions on time scales of order the Planck time, but with rules that are changing on a much longer time scale.  During periods much shorter than the age of the universe but much longer than the Planck time, we are still dealing with a fixed permutation $S$, 
 but 't Hooft hypothesizes that the effective state of the universe induced by the energy cutoff/RG procedure is related to the ontological basis by a non-trivial unitary.  This is equivalent to saying that the effective operators in the RG reduced theory are not subject to the $U(1)^N$ gauge constraint.  It converts $DOQM_N$ to $OQM_P$ for some $P < N$.  There remain many issues to be worked out regarding the necessity for an era of adiabatic evolution at scales higher than those of the standard model, which allows for the definition of an approximately conserved energy, and can then be used to define the RG procedure.

 \section{$S_N$ dynamics without $U(1)^N$ Gauge Invariance}
 
 If we drop the requirement of $U(1)^N$ gauge invariance, we obtain the theory we call $OQM_N$. It is just an ordinary quantum theory with a special form of evolution operator.  Operators not diagonal in the RG reduced ontic basis are physical and probabilities for histories of eigenvalues of these operators do not satisfy the PSR.  
 As a consequence, the "classical" evolution of the effective ontological variables {\it does not determine the evolution of all of the variables that are presumed observable in the OQM formalism.}  We have to ask why we can not supplement the ontological permutation evolution with an arbitrary unitary in the group that multiplies individual ontological basis states with independent phases, without changing the evolution law of the ontic variables.   Somehow, if one wants to imagine that $OQM_P$ came from $DOQM_N$, then the microscopic gauge invariance, which is broken by the energy cutoff, must fix these phases.

 In OQM, as opposed to DOQM the sum over histories rule is satisfied for {\it amplitudes} of histories but not for probabilities of histories of the eigenvalues of operators not-diagonal in the ontic basis.  These operators are declared unphysical by the gauge invariance of the deterministic theory. Within this  paradigm, it seems natural to consider the possibility that our system lives in $N + 1$ dimensions, with evolution restricted to $S_{N + 1}$, but assume that the effective operators we measure have very small matrix elements off-diagonal between the singlet state and the $N$ dimensional subspace.  For large $N$, the space of operators which connect the two subspaces, has dimension much smaller than the space of those which preserve their separation.  This is effectively the posture taken in\cite{finite}.  
 
 The general element of $S_{N + 1}$ is generated by $S_N$ transformations on a particular basis, and the Fourier transform operator $F$.  Since $F$ squares to a reflection in the "clock" of a particular basis, which is in $S_N$, we can write the general element as \begin{equation} s_U s_V , \end{equation} where $s_U$ is a permutation of the $U$ basis and $s_V$ an independent permutation of the conjugate $V$ basis.  
 
 A general element of the permutation group is a product of commuting cycles of prime length.  Thus, the eigenvalues of a given permutation $s_U$ have the form
 \begin{equation} e^{i\phi} = e^{2\pi i \sum_j \frac{k_j}{p_j} } , \end{equation} where $k_j$ runs from $1$ to $p_j$ and the $p_j$ are primes satisfying $\sum p_j = N$. it is known that for large $N$ the maximal order of an element of $S_N$ behaves like $e^{\sqrt{N {\rm ln}\ N}}$\cite{largeorder}. Since there are only $N$ distinct eigenvalues, the spectrum of these large order elements is rather sparse in the space of all $ e^{\sqrt{N {\rm ln}\ N}}$-th roots of unity.  
 
 It is obvious that $s_U$ is diagonal in the $V$ basis, and $s_V$ in the $U$ basis, so any given OQM evolution in $N + 1$ dimensions will have the Trotter form in the ontological basis of $N$ dimensional space and its Fourier transform.  Thus, the question arises of how well we can approximate the spectrum of a general unitary in $SU(N)$
 with that of an element of $S_N$.  We've seen that the spectrum of generic $S_N$ elements is quite chaotic, and it certainly has energy differences small enough to describe evolution over time scales $\gg {\rm ln}^q\ N$, for any fixed $q$.  This means that, at least for the purposes of evolution over times short compared to a power of the maximal entropy of the system, the operators $s_{V,U}$ can be quite general.  Combining this with the product form of the short time evolution operator for all known interesting models\footnote{We're of course assuming an approximation to the interesting model with a finite number of states.}, we conclude that Kornyak's $SU(N + 1)$ quantum mechanics is an adequate framework for any $N$ dimensional system over time scales at least as long as powers of the entropy.
 Of course, if we allow the full group ${\cal G}_{\cal F} (N)$ instead of its $SU(N + 1)$ subgroup then we can certainly approximate any Trotter evolution with arbitrary accuracy, by increasing the size of ${\cal F}$.
 
 A grandiose way of describing Kornyak's QM is to say that classical mechanics with $N$ states, has a quantum mechanical conservation law, the character of the $S_N$ irrep, and the choice of the value of that conservation law can generate an adequate approximation to any known quantum theory with $N - 1$ states for times much shorter than $N$.  The elementary unit of time is of course assumed to be much shorter than any time probed by experiment, most plausibly the Planck time.  If the restriction to finite $N$ comes about because we live in an asymptotically de Sitter space, then the lifetime of any localized object, from the point of view of {\it any} static measuring apparatus, is indeed $\ll N$.
 
 The considerations above ignore an interesting unsolved question.  Does the spectrum of the Hamiltonian completely characterize a quantum mechanical system? This is related to questions like, "Can one hear the shape of a drum?" or "Does the scattering matrix determine the potential?", or to the bootstrap program in conformal field theory.  These problems are notoriously difficult and have not received definitive answers despite decades of work by talented people.  Similarly, with the exception of free $1 + 1$ dimensional field theories, 't Hooft has not succeeded in exhibiting an OPM approximation to local field theory.  We do not resolve these issues here, but will address one simple question that is related to locality.  We'll also note that in lattice approximations to field theory, the Trotter form of the evolution operator is sufficient to isolate the local operators.
 
 It is obvious that OQM for a two state system cannot approximate the predictions of quantum mechanics within experimental tolerance, although ${\cal G}_{\cal F} (N)$ quantum mechanics can.  So how can OQM account for the experimental results on two state systems?  Only by embedding the system in a much larger system. Page's theorem\cite{page} tells us that a generic pure state of the large system will have an almost maximally uncertain reduced density matrix for the two state system, so in order to explain quantum interference experiments on the two state system, we must include a quantum description of the "experimental apparatus" that measures the properties of the two state system.  The quantum mechanical notion of measurement is maximal entanglement of the two state system with macroscopic collective coordinates, "pointers", of the measurement apparatus.  Quantum probabilities for histories of the pointers satisfy the Probability Sum Rule (PSR), which enables us to apply Bayes' rule for conditional probabilities, up to corrections exponential in the number of atoms in the pointer.  The monogamy of entanglement tells us that we can ignore the rest of the universe in our discussion of the two state plus apparatus subsystem.
 The possibility of maintaining the isolation of that subsystem for the duration of experiments depends on many details of the theory of the real world.  Most important is the approximate description of interactions by local field theory, but also the fact that the amplitudes for emission of soft massless particles are all small, and do not affect the quantum state of the pointer variables.  
 
 A generic state of the q-bit plus apparatus system has the q-bit maximally entangled with the apparatus.  The essence of the notion of measurement is that that entanglement can evolve to be entanglement of, say the $\sigma_3$ eigenstates of the q-bit with macroscopically different values of the pointer variables.  We then use Bayes' rule for pointer probabilities to make future predictions based on the assumption that the pointer was in the state entangled with that in which the q-bit was in a particular $\sigma_3$ eigenstate. This is the procedure colloquially called {\it collapse of the wave function}.  An experiment proceeds by decoupling the apparatus from the q-bit, usually by manipulating its translational collective coordinate and again relying on locality and decoupling of soft emission.  The q-bit then evolves on its own, after which we recouple it to the apparatus and make a measurement.  
 
 Now assume the full q-bit plus apparatus evolves by discrete $S_{N + 1}$ dynamics in its $N$ dimensional Hilbert space, with a discrete time interval much shorter than the time resolution involved in the experiment.   What is the effective evolution operator for the q-bit during the course of the experiment?  We know of many examples of ordinary quantum systems for which the answer to this question is clear. The single q-bit can evolve over the course of the experiment via a possibly time dependent Hamiltonian $B_a (t)\sigma_a$ acting only on the q-bit.  The functions $B_a (t)$ are determined by the classical evolution of the collective coordinates of the apparatus, which we chose in our application of Bayes' rule to the initial wave function.   In other words, $S_{N + 1}$ evolution of a large system can approximate ordinary QM for small subsystems if it can approximate the predictions of QM for large local systems.   As we've emphasized, for experiments that can actually be performed over many times the lifetime of our universe, this question is equivalent to the question of whether the complete spectrum of a cutoff quantum field theory is enough information to reconstruct the local operators of the theory.  If we want to approximate QM with arbitrary mathematical precision, more than could be tested in conceivable experiments, then we have to generalize $OPM$ to allow for ${\cal G}_{\cal F} (N)$ evolution with arbitrarily large ${\cal F}$.
 
 In summary, any model that can be viewed as the limit of a perturbation of a quadratic Hamiltonian for bosonic or fermionic canonical variables has an evolution operator which, after truncation to an $N$ dimensional Hilbert space, has the form
 \begin{equation} f(U) g(V) , \end{equation} where $U,V$ is a particular pair of conjugate clock and shift operators\footnote{All such pairs are unitarily equivalent.} over short enough time scales. $f,g$ are unitary operators in the commutants of $U$ and $V$ respectively. 
 We argued that any such evolution could be well approximated by evolution in the $N$ dimensional representation of $S_{N + 1}$, the finite QM of\cite{finite}, as long as the phase differences between eigenvalues of both $f$ and $g$ were $\geq q(N)$ with $q(N) \leq 1/N$.  We did not establish the precise value of $q(N)$ because, for any modestly macroscopic system, the time required to measure phase differences of order $1/N$ is exponentially longer than the age of the universe.  If we require more mathematical precision of our approximation than can be tested in conceivable experiments, then we must expand the finite group to ${\cal G}_{\cal F} (N)$, the semi-direct product of $S_{N + 1}$ and individual ${\cal F}$ phases on each element of the eigenbases of $U$ and $V$.  Even in an eternal universe with potentially infinite resources, no finite sequence of experiments could rule out the possibility that QM was supplemented by a condition restricting time evolution to be discrete and restricted to ${\cal G}_{\cal F} (N)$ for some large enough ${\cal F}$.
 
 A question we left open was whether the spectrum of the Hamiltonian of a quantum system was sufficient to reconstruct spatially localized operators in the theory.  All of our notions of subsystems and measurement, rely implicitly on an approximate notion of spatial locality.  For lattice field theories that are perturbations of a quadratic Hamiltonian, the answer to this question is obviously yes.  Specification of the Hamiltonian is tantamount to specification of the operators $s_{V,U}$.  In a lattice field theory, either the $U$ or $V$ basis is one in which the number operators for fixed momentum are all diagonal, and this is enough information to construct all local lattice operators.

  \section{Quantum Evolution and Random Unitaries} 
 
 As mentioned in the introduction, there is lots of evidence that the eigenvalue distributions of the Hamiltonian of a many body chaotic quantum system are well modeled by Random Matrix Theory (RMT). This has led some researchers to conjecture that the long time evolution of chaotic quantum systems is described by evolution by an operator chosen randomly from the Haar distribution on the unitary group.  The results of this paper show that this conjecture is false for time independent Hamiltonians.  Indeed, for any fixed Hamiltonian, the probabilities to be in each of its energy eigenstates are conserved.  These are the "ontological conservation laws" for a time independent system.  This means that quantum recurrences, where an initial state comes back to itself with arbitrary accuracy (with some metric on the space of density matrices), can occur, but that the set of initial states falls into classes which cannot evolve into each other, no matter how long a time passes.
 
 Computational complexity is defined for a system with $2^B$ states, with respect to a computational basis that diagonalizes a set of $B$ Pauli operators $Z(i)$.  One then defines a set of "gates", unitary transformations that act only in one, or a pair of tensor factor Hilbert spaces.   With appropriate choice of these gates (one needs two single q-bit and one two q-bit gates) one can prove the analog of the classical Church-Turing theorem.  Namely, any unitary $U$ on the Hilbert space can be approximated\footnote{Usually, in the sense of the distance function $ || (U - U_n) ||_1 \equiv {\rm Tr}\ \sqrt{(U - U_n )^{\dagger} (U - U_n )} < \epsilon$.} by a sequence of $n$ of these gates where
 \begin{equation}  2^{2B} {\rm ln}\ (\epsilon^{-1}) < n <  2^{2B} {\rm ln}^c\ (\epsilon^{-1})\ \ \ \ 1\leq c \leq 2 . \end{equation}  Here $U$ is a random unitary and $U_n$ is the product of the sequence of $n$ gate operations.  
 Note, this does not look like a typical Hamiltonian evolution, and it certainly is not time independent, which would require $$ U_n = (U_{k})^{n/k} , $$ where $k \sim 1$.  
 
 A general quantum circuit, a product of an arbitrary sequence of the elementary gates, does not resemble even a conventional time dependent Hamiltonian system because each gate acts on at most two q-bits. Thus a single discrete time evolution by a Hamiltonian that is a sum over terms that act on a small number of commuting $1$ and $2$ q-bit operators.  Such a Hamiltonian is, in the language of computer science, a massively parallel computer, which implements a number of gates proportional to the number of q-bits, simultaneously.  Thus, the discrete time of a typical many-body Hamiltonian is the {\it depth} of a parallel quantum circuit.  
 
 The question we want to ask then is whether a discrete time independent quantum evolution $e^{ -i n H}$ of a many body system can approximate a general unitary transformation $W$.   For simplicity we will work with the $L_2 $ norm of the operator  difference 
 \begin{equation} ||e^{-inH} - W||_2  \equiv \sqrt{{\rm tr}\ M^{\dagger} M} , \end{equation} where $M = e^{- inH} - W$.  We work in the basis where $H$ is diagonal 
 \begin{equation} ||e^{-inH} - W||_2^2 = \sum_{i,j
 } (e^{in \phi_i} \delta_{ij} - w_{ji}^*) (e^{-i n \phi_i} \delta_{ij} - w_{ji})
 ) \end{equation}
 \begin{equation} = \sum_i (2 - e^{in\phi_i} w_{ii} - e^{-in\phi_i} w_{ii}^*)  . \end{equation} 
 If the phases $e^{i\phi_i}$ are all $p$-th roots of unity, and $p$ is large enough, then we can make the real part of $e^{in\phi_i} w_{ii} $ positive in order to make this norm as small as possible, but the absolute values of all the $w_{ii}$ can be small.  For example, if $W$ is diagonal in the Fourier transformed basis to that in which $H$ is diagonal, then $|w_{ii}| = N^{-1/2}$.  The square of the $L_2$ norm is thus $2(N - \sqrt{N})$.  Thus, time independent evolution cannot achieve maximum complexity.  
 
 It is clear that in a certain sense, this conclusion is old news.  In a time independent system, energy eigenstates evolve into themselves.  In quantum mechanics with a finite number of states, once we have fixed the eigenvalues of all mutually commuting operators that also commute with the Hamiltonian, we expect a non-degenerate energy spectrum.
 The relevant result for statistical mechanics is the {\it eigenstate thermalization hypothesis}\cite{eth} (ETH).  If an eigenstate belongs to the dense part of the spectrum of a non-integrable quantum system, then the density matrix for small enough subsystems will be thermal w.r.t. the subsystem energy even if the whole system is in an exact energy eigenstate.  However, if we take a small band of energies, of size $\Delta E$ and a generic superposition of states from this band, then one occasionally assumes that the expectation values of products of Heisenberg operators in that state will approach their thermal expectation values at a temperature of order $\Delta E$.  This is true for a special class of operators, whose matrix elements between eigenstates follow the ETH rules, but the thermal density matrix and the true density matrix will have an $L_1$ distance
 \begin{equation} || e^{- H/T} - \rho ||_1 = \sum_i | e^{- E_i / T} - p_i | , \end{equation} which is definitely different from zero and has no reason to be small.  For example, in a dense band of states we could take an initial state with $p_i = 0$ for a randomly chosen set of half or one quarter of the states.
 
 Ergodic averaging, defining $\bar{\rho} = \tau^{-1} \int_0^{\tau} dt\ \rho (t) $, does not change this conclusion.  It plausibly eliminates off diagonal elements of the density matrix in the energy basis, but does not change the $p_i$.  This rather elementary observation has serious implications for our understanding of how statistical mechanics  emerges from an underlying mechanical theory.  The long time, or time averaged density matrix for a pure state under chaotic\footnote{The meaning of chaotic in quantum mechanics is not obvious. We are using it in the sense that the system is large and its Hamiltonian has, to a good approximation, the eigenvalue statistics of a random matrix ensemble (RME).  it is possible that one only need require the first few moments of the eigenvalue distribution agree with an RME.} quantum time evolution {\it does not} approach that of a thermal ensemble, in any of the conventional $L_p$ metrics on the space of density matrices.
 
 A rather different approach to the derivation of statistical mechanics goes back at least to the work of Khinchin\cite{khinchine}.  That is, one only requires that expectation values of a limited class of operators approach their thermal averages.   This is the approach used in the ETH\cite{eth}.  Recent work by Lucas and the present author\cite{tbal} sheds some light on this approach.  These authors considered a quantum system whose Hilbert space was a tensor product of finite dimensional Hilbert spaces defined on the points of a graph.   The Hamiltonian was short ranged in the sense that interactions between points that were distant from each other on the graph fell off at least exponentially in the distance.  As a consequence, one could define large subsystems containing $V\gg 1$ points such that, up to exponentially small corrections, each subsystem interacted only with a small number of "nearest neighbors".  That is, the Hamiltonian has a decomposition 
 \begin{equation} H = \sum_X H(X) + \sum_{X \sim Y} H(X,Y) , \end{equation} with $|| H(X) || > V^{1/d} || H(X,Y) || $, where the vertical bars refer to the maximal eigenvalue.  The sums are over large subsystems and the $\sim$ refers to the nearest neighbor restriction.  Define $L = V^{1/d}$ .   The individual Hamiltonians $H(X)$ act on Hilbert spaces of dimension $e^{c V}$ with $c \geq {\rm ln}\ 2$, and have typical eigenvalues of size $V \epsilon (X)$.  They are mutually commuting and we assume the full Hilbert space is exactly spanned by their non-degenerate\footnote{Degeneracies indicate symmetries: operators that commute with the Hamiltonian.  Local symmetry operators can be diagonalized, and if their spectrum is dense lead to additional hydrodynamic variables. Symmetries associated with space translation and/or higher $p$ form charges introduce  new features, which have not yet been analyzed completely.} spectra.   One can then show that the diagonal elements of the density matrix in the $\epsilon (X)$ basis satisfy a Fokker-Planck equation and define a stochastic hydrodynamic flow of energy.  The hydrodynamic time scale is $\sim L^{-2}$.  Central to the derivation is the fact that the spectra $\epsilon (X)$ have approximate degeneracies, with splittings as small as $e^{- V}$ in fundamental energy density units.  These degeneracies are unresolved on the hydrodynamic time scale and provide a definition of the local entropy $S(\epsilon (X))$, which satisfies the usual postulates of hydrodynamics.  
 
 If, in some interval of energy density,  the entropy\footnote{The entropy of course depends on the choice of initial state.  The statements in the text are valid when the initial state has reasonable overlap with a large collection of states in the interval.} grows like a power $\epsilon^p (X) $ with $1 > p > 0$, so that the specific heat is positive, then the regulated density of states $e^{S(\epsilon (X))} e^{- \beta V \epsilon (X)}$ has a sharp maximum in the interval, for some value $\beta$.  Operators that are insensitive to the details of the quasi-degenerate states contributing to the entropy (this is the import of the smoothness assumption of ETH) will have Boltzmann ensemble expectation values and Gaussian fluctuations, to leading order in $L$.  In our example where $1/2$ the eigenstates in the interval have zero weight in the initial quantum state, this will be irrelevant for operators whose matrix elements do not distinguish between the different eigenstates.  If the missing half is determined by a criterion like "every other eigenstate" then this is clearly the case for many operators, including all those that are actually easy to measure.  If it is "the lowest half" this will simply lead to a shift in the temperature.   There might be interesting cases intermediate between these two extremes, which would not match the Boltzmann ensemble for simple operators.
 
 In summary, statistical mechanics does not depend on ergodic hypotheses invoking random unitary dynamics even for a time averaged density matrix.  Time independent quantum systems do not have random unitary dynamics, even though their eigenvalue spectrum matches that of a random unitary.

 \section{Conclusions}
 
 Finite quantum mechanics replaces the unitary group $U(N)$ by the group ${\cal G}_{\cal F} (N)$ generated by permutations of the elements of a single basis, ${\cal F}$ phases on any one basis element, and the Fourier transform operator.  It is the semi-direct product of the action of $S_{N + 1}$ on $N$ dimensional Hilbert space, and ${\cal F}^{2N}$.  Although it cannot mimic an arbitrary unitary evolution with arbitrary precision as ${\cal F}$ gets large it can approximate any model whose evolution over short enough time periods is of the form $f(U) g(V)$ where $U,V$ are a canonical clock/shift operator pair.  We call this class of models a Trotter equivalence class.  A Trotter equivalence class includes any finite perturbation of a free lattice field theory of bosons\footnote{For bosons we include the replacement of the usual canonical variables by a finite dimensional clock/shift pair in our definition of the lattice approximation.} or fermions, and, {\it a fortiori} any conformal field theory that arises as a critical point of such a lattice model.  Among all models studied by theoretical physicists, this leaves only a few superconformal field theories in $4,5$ or $6$ dimensions that cannot obviously be approximated by finite quantum mechanics.   If the world is modeled by a system in a fixed Trotter equivalence class, experiment will never be able to distinguish ${\cal G}_{\cal F}$ QM from $QM$ if ${\cal F}$ is chosen large enough.
 
 We also studied 't Hooft's $DOQM_N$, and its ungauged version $OQM_N$, which allows superpositions of quantum states, but insists that the dynamics be restricted to the $S_N$ subgroup of $U(N)$. Within this context it makes sense to study the $S_{N + 1}$ subgroup instead, because $S_{N + 1}$ dynamics has a quantum mechanical conservation law, the projector on the $S_{N + 1}$ invariant state, so although the dynamics exhibits quantum interference between histories of the ontological states of the $N$ dimensional system, it can be viewed as satisfying the sum over histories rule for probabilities of ontic states in $N + 1$ dimensions.  This is a kind of "hidden variable" theory.   One assumes an underlying "classical" dynamics of some particular basis, but allows for the possibility that "our measurements" couple to operators non-diagonal in that basis. 
 
 't Hooft's cellular automaton interpretation of QM imposes a $U(1)^{N + 1}$ gauge invariance on $OQM_{N + 1}$, but violates that gauge invariance via a cutoff on the spectrum of the evolution operator.  The evolution eigenvalue basis is the Fourier transform of the ontic basis. We argued that this procedure could only reproduce the real world if the underlying dynamics was time dependent, allowing the classical evolution of ontic variables to range over more than a single cyclic perturbation, and the microscopic energy spectrum to be chaotic.  The permutation $S(t)$ corresponding to the microscopic evolution of the universe must become adiabatic\footnote{Really only the product of the $S(t_i)$ over all the Planck intervals in the time scale $E^{-1}$ must become approximately independent of {\it which} interval is under consideration.  Note that adiabaticity is necessary even to discuss the concept of an energy cutoff.} during the period of the universe we observe.  't Hooft's hypothesis is that the effective evolution operator in the energy cutoff subspace is still a permutation, but that the gauge invariance is no longer imposed on the effective states.  That is, in the language of this paper, "time dependent $DOQM_N$ in an adiabatic regime and after an RG transformation to cut off the energy spectrum, is equivalent to $OQM_M$ for some $M < N$".  The present author does not understand 't Hooft's argument, that the actual values of collective coordinates of macroscopic objects, are simply functions of the the underlying ontological variables.  If the RG in quantum energy space is invoked in order to understand why the appropriate variables to describe particle physics are "effectively quantum mechanical", then why should the collective coordinates, conventionally constructed as averages of quantum fields over large volumes, not be quantum mechanical as well?
 
  It does not seem unreasonable, especially for the extremely large $N$ characterizing any remotely macroscopic system, instead make the simple postulate of Kornyak.  A slightly weaker form of that postulate is that matrix elements of "our operators" between the two irreducible representations of $S_{N + 1}$ in $N + 1$ dimensional space are small.    Within $OQM_{N + 1}$, with this assumption we can approximate all models within a Trotter equivalence class by OQM as long as the time interval over which the two systems have to match is much shorter  than $N$ in fundamental time units.   If the fundamental unit is the Planck time, $\sim 10^{-44}$ seconds, then the current age of the universe is $10^{61}$ which, would saturate this bound for $200$ q-bits.   Since even quite small chunks of matter have ${\rm ln}\ N \sim 10^{20}$ it is clear that the correct model of our world, at least as far as the spectrum of the Hamiltonian is concerned, could easily be fit by the ungauged OQM.
 
 A question that remains is the extent to which the spectrum of the Hamiltonian is enough to reproduce the correlation functions of {\it e.g.} a local field theory.  Some insight into this question comes from the study of integrable field theories in $1 + 1$ dimensions, where the exact spectrum and S matrix have been known for decades, but the construction of local operator correlation functions in all but the simplest models has not been carried out.   A survey of the state of the art can be found in\cite{integrable}. From this reference one learns that the form factor approach to the construction of correlators has not even been proven to converge.
 
 For continuum quantum field theories, the question of whether the spectrum of the Hamiltonian allows one to reconstruct correlation functions is equivalent to the question of whether a list of primary operator dimensions is enough to determine all of the operator product expansion coefficients, though there might also be delicate convergence questions involved.  This problem could be approached by conformal bootstrap techniques.
 
 Of course, for lattice approximations to field theory, the Trotter equivalence class already identifies the "elementary" local fields, so the question at hand is merely the conventional one of extracting local fields as limits of lattice variables.

 \section{Appendix: Philosophical Questions}
 
 There are two fundamentally different views of the systems of equations that we call The Laws of Physics.  The first view takes them to be a scale model of "what is going on in the real world".  An exemplar of this view is the relation between a planetarium and the solar system.   Since everything in the real world "definitely happens", the only reason for the concept of probability to be brought into a description of physics is "human"\footnote{I use "human" as a shorthand for "information gathering and analyzing system".} frailty.  We are trying to make predictions about the future course of the world we observe but are unable to gather all of the initial data necessary to make those predictions, and/or make sufficiently accurate measurements to verify/falsify those predictions with $100\%$ confidence.   
 
 Mathematically, this point of view implies linear equations for the time evolution of probability, even if we decide that some of the variables necessary to predict the future are forever hidden from our view.  If the system has only a finite number of states, and the dynamics is reversible, then time evolution is a permutation of the states, which might be time dependent.  The PSR then follows as a mathematical axiom.  This is the mathematical statement of the philosophical concept of determinism.  When I first learned of Koopman's Hilbert space formulation of QM as an undergraduate, I believed that it showed that determinism was an illusion because that formulation has quantum interference for amplitudes involving operators that do not commute with the phase space coordinates.  I realized only later that determinism could be imposed mathematically by insisting that changes of phase of the wave function on phase space were gauge transformations, so that all such operators were non-gauge invariant.  The analog of this for finite dimensional systems is the difference between $OQM$ and $DOQM$.
 
 The second view of the laws of physics takes them to be prediction algorithms, rather than scale models of "what is actually going on".  The universe does what it is doing, independent of our equations, and each event is actually unique.  Our laws are just tools to help us understand what's going to happen before it happens.  From this point of view there is nothing objectionable to saying that actually the fundamental laws are statistical, even for a hypothetically perfect "human" observer.  
 
 Quantum mechanics is a natural and inevitable statistical theory of prediction.  A list of data specifying the state of a system can always be thought of as a vector in a vector space, and functions of that data are diagonal operators in a particular basis for that space.  There is a canonical complex\footnote{Complex numbers are introduced for the usual reason: to allow for the solution of arbitrary algebraic equations.} scalar product 
 on any such space and the $N$ dimensional complex Pythagorean theorem shows us that every unit vector in the space defines a mathematical probability distribution on the space of all unit vectors, or equivalently on the spectra of {\it all} normal operators.
 Classical notions of dynamics restrict attention to the $S_N$ subgroup of all unitary operators acting on the space, and preserving a particular "ontological" basis.  Kornyak's observation is that the mathematically natural decomposition of the space into irreducible $S_N$ representations, automatically forces us to consider both an ontological basis of the $N - 1$ dimensional irrep and its canonical conjugate basis, in which cyclic shifts of period $N - 1$ of the ontological basis are diagonal.  The considerations of the present paper show that Kornyak's formalism can accurately reproduce all of the results of conventional quantum mechanics for all models whose short time evolution operator is the product of an operator diagonal in the ontological basis and and operator diagonal in the conjugate basis, as long as the time over which we require the approximation to work is $\ll N$.  Since any finite dimensional approximation to models that have been studied in the literature is of the Trotter form, and since ${\rm ln}\ N$ is $ > 10^{15}$ for any "macroscopic" subsystem of the world, this is enough to do theoretical physics.   If we want to approximate quantum systems with arbitrarily long recurrence times, we must adjoin ${\cal F}$ factors to the group of allowed time evolution operators.
 
 In quantum field theory or any discretization of it, the classical behavior of the collective coordinates of macroscopic objects is completely accounted for by their construction as averages of quantum fields over  volumes $V$ large in microscopic units.  The fluctuations of such operators are of size $V^{- 1/2}$ and their probability distributions obey linear classical statistical equations, up to corrections of order $e^{-V}$. These are irrefutable mathematical results, and give a complete understanding of the apparently deterministic behavior of macroscopic objects, as well as the deviations from that behavior seen in phenomena like Brownian motion.  
 
 At the moment, I do not see any principled scientific procedure, which could experimentally distinguish an explanation for determinism of the macro-world as a direct manifestation of determinism of the ultimate microscopic description of physics, from the decoherence arguments referred to in the paragraph above.  The only hope I can see for that is a much clearer mathematical argument showing that microscopic determinism plus cutoffs on the spectrum of the microscopic evolution operator could reproduce the successes of QM on scales between $10^{-30}$ seconds and $10^{-13}$ seconds, but avoid the explanation of macroscopic determinism as an emergent property of large bound states of quantum particles.  
 
 By formulating determinism as a kind of gauge invariance, and 't Hooft's attempt to reproduce quantum behavior by invoking a gauge violating cutoff scheme/renormalization group, I hope I've made some contribution to the resolution of this question.
 
 My own prejudice is to go back to my undergraduate point of view:  Koopman showed that even classical mechanics had a hidden quantum theory underlying it.  The work of Kornyak shows that the mathematical fact that discrete classical mechanics has a quantum conservation law (the projection on the $S_N$ singlet subspace of the Hilbert space), naturally generates a set of truly quantum systems, which can encompass finite dimensional approximations to all known models of theoretical physics.

 \vskip.3in
\begin{center}

{\bf Acknowledgments }\\
This work was motivated by questions that my son, Eric Banks, asked me some years ago. I'm also grateful to G. 't Hooft for commenting on an early version of the manuscript and pointing out some errors. The work was supported in part by the Department of Energy, under grant DE-SC0010008. 
\end{center}

\end{document}